\documentstyle[epsfig,preprint,aps]{revtex}           
\begin{document}       
\tighten
           
\def\bea{\begin{eqnarray}}          
\def\eea{\end{eqnarray}}          
\def\beas{\begin{eqnarray*}}          
\def\eeas{\end{eqnarray*}}          
\def\nn{\nonumber}          
\def\ni{\noindent}          
\def\G{\Gamma}          
\def\d{\delta}          
\def\l{\lambda}          
\def\g{\gamma}          
\def\m{\mu}          
\def\n{\nu}          
\def\s{\sigma}          
\def\tt{\theta}          
\def\b{\beta}          
\def\a{\alpha}          
\def\f{\phi}          
\def\fh{\phi}          
\def\y{\psi}          
\def\z{\zeta}          
\def\p{\pi}          
\def\e{\epsilon}          
\def\ve{\varepsilon}          
\def\cl{{\cal L}}          
\def\cv{{\cal V}}          
\def\cz{{\cal Z}}          
\def\pl{\partial}          
\def\ov{\over}          
\def\~{\tilde}          
\def\rar{\rightarrow}          
\def\lar{\leftarrow}          
\def\lrar{\leftrightarrow}          
\def\rra{\longrightarrow}          
\def\lla{\longleftarrow}          
\def\8{\infty}          
\newcommand{\fr}{\frac}

\title{Calculation of a Class of Three-Loop Vacuum Diagrams       
with Two Different Mass Values}

\author{J.~-M. Chung\footnote                  
  {Electronic address: chung@ctpa03.mit.edu}}
\address{Center for Theoretical Physics,
Massachusetts Institute of Technology,\\
Cambridge, Massachusetts 02139}

\author{and\\
~\\B.~K. Chung\footnote                  
  {Electronic address: bkchung@nms.kyunghee.ac.kr}}               
\address{ Research Institute for Basic Sciences                   
and Department of Physics\\ Kyung Hee University, Seoul  130-701, Korea\\
{~}} 
       
\date{MIT-CTP-2802,~~~~ November 1998}                  
\maketitle              
\draft              
\begin{abstract}           
\indent           
Using the method of Chetyrkin, Misiak, and M\"{u}nz
we calculate analytically a class of three-loop vacuum diagrams with two           
different mass values, one of which is one-third as large as the other.           
In particular, this specific mass ratio is of great interest in relation to 
the three-loop effective potential of the $O(N)$ $\f^4$  theory. All pole 
terms in $\e=4-D$ ($D$ being the space-time dimensions in a dimensional
regularization scheme) plus finite terms containing the logarithm of mass
are kept in our calculation of each diagram. It is shown that 
three-loop effective potential
calculated using three-loop integrals obtained in this paper agrees, 
in the large-$N$ limit, with the overlap part of leading-order 
(in the large-$N$ limit) calculation of Coleman, Jackiw, and Politzer 
[Phys. Rev. D {\bf 10}, 2491 (1974)].

\end{abstract}              
                   
\pacs{PACS number(s): 11.10.Gh}           
\section{Introduction}             
Although quantum field theory has a long history and there are      
a number of different approaches to it,      
Feynman diagrams are still the main source of its dynamical      
information. The necessity to know more exactly about the characteristics      
of specific physical processes and corresponding quantum-field-theoretic      
functions themselves      
stimulates the calculation of radiative corrections of ever higher order.      
The main purpose of this work is to calculate a class of three-loop diagrams 
shown in Figs.~1 to 4 in a dimensional regularization scheme \cite{tv}.       
Most of the diagrams carry two kinds of propagator lines:       
A-type lines and B-type lines. The mass parameter of the B-type line is       
one-third as large as that of the A-type line. These diagrams are interesting       
because they are genuine three-loop integrals --- genuine in the sense that       
they cannot be factorized into lower-loop integrals --- appearing in the 
three-loop effective potential of the massless $O(N)$ $\f^4$ theory. 

The calculation of two-loop effective      
potential of the same theory can be found in Jackiw's classic paper      
\cite{jk}. 
Very recently, the three-loop effective potential was calculated using
three-loop integrals obtained in this paper: Details of calculation
itself of three-loop integrals are described here and details of effective 
potential calculation itself were reported elsewhere \cite{jkps}.
Our final expression of three-loop effective potential in Ref.~\cite{jkps}
is appended to the end of the summary section of this paper in order to 
compare the result of Coleman, Jackiw, and Politzer \cite{cjp}, since this 
comparison was not done in Ref.~\cite{jkps}. 
The diagrams $J(a)$ in Fig.~1, $K(a)$ in Fig.~2, $L(a)$ in Fig.~3, 
and $M(a)$ in Fig.~4 were already calculated in the        
literature \cite{jm,cmm,pv}. The calculation of these diagrams are reproduced        
in this paper using the method of Chetyrkin, Misiak, and M\"{u}nz \cite{cmm},       
with which we are also going to calculate other diagrams.       
          
The organization of this paper is as follows: In Sec.~II we decompose       
each three-loop integral in Figs.~1 to 3 into a three-loop integral yet      
to be calculated and the known two- and one-loop integrals      
by separating the pole part and the finite part of the two-loop integral       
inside each three-loop integral, and discuss the calculation of diagrams       
in Fig.~4. Decomposed three-loop parts are calculated in Sec.~III. We       
summarize our results of three-loop calculations by listing them in the      
$\e$-expanded form and simply give the result of the three-loop
effective potential of massless $O(N)$ $\f^4$ theory in order to compare it 
with earlier result obtained in the large-$N$ limit \cite{cjp} in Sec.~IV. 
In the Appendix,       
we list one- and two-loop integrals which are needed        
in the calculation of three-loop integrals with the method of Chetyrkin,       
Misiak, and M\"{u}nz. 

\section{Preliminary: Decomposition of Three-Loop Integrals}                 
It is convenient to use the Euclidean metric in our discussion.        
Thus throughout the paper the momenta appearing in the formulas are           
all (Wick-rotated) Euclidean ones and the abbreviated integration measure is             
defined as              
\beas              
\int_k=\m^{4-D}\int{d^D k\ov (2\p)^D}\;,              
\eeas              
where $D=4-\e$ is the space-time dimension in the framework of dimensional             
regularization \cite{tv} and $\m$ is an arbitrary constant with mass            
dimension.           
          
First let us define ${\cal{P}}_n$ as follows:          
\bea          
{\cal{P}}_n&\equiv&                 
\int_{kpq}{1\ov (k^2+\s^2)^n(p^2+\s^2)[(p+k)^2+\s^2]          
(q^2+\s^2)[(q+k)^2+\s^2]}\;.\nn          
\eea          
Then with the following separation of one-loop integral into       
the pole part and the finite part          
\bea          
\int_p{1\ov (p^2+\s^2)[(p+k)^2+\s^2]}&=&          
{1\ov (4\p)^2}\biggl({\s^2\ov 4\p\m^2}\biggr)^{\!\!\!{-\e/2}}               
\biggl[{2\ov \e}+F(k^2)\biggr]\;,\label{dcf}          
\eea          
we readily see that three-loop integrals $J(a)$ in Fig.~1, $K(a)$ in Fig.~2,
and $L(a)$ in Fig.~3 are given as          
\bea               
J(a)&\equiv&{\cal{P}}_0=                
{1\ov (4\p)^4}\biggl({\s^2\ov 4\p\m^2}\biggr)^{\!\!\!{-\e}}               
\int_k \Bigl(F(k^2)\Bigr)^{\!2}+{4W_1\ov (4\p)^2\e}               
\biggl({\s^2\ov 4\p\m^2}\biggr)^{\!\!\!{-\e/2}}\;,\nn\\           
K(a)&\equiv&{\cal{P}}_1=          
{1\ov (4\p)^4}\biggl({\s^2\ov 4\p\m^2}\biggr)^{\!\!\!{-\e}}               
\int_k{[F(k^2)]^2\ov k^2+\s^2}               
+{4W_4\ov (4\p)^2\e}\biggl({\s^2\ov 4\p\m^2}\biggr)^{\!\!\!{-\e/2}}               
-{4S_1\ov (4\p)^4\e^2}\biggl({\s^2\ov 4\p\m^2}           
\biggr)^{\!\!\!{-\e}}\;,\nn\\              
L(a)&\equiv&{\cal{P}}_2=                
{1\ov (4\p)^4}\biggl({\s^2\ov 4\p\m^2}\biggr)^{\!\!\!{-\e}}               
\int_k{[F(k^2)]^2\ov (k^2+\s^2)^2}               
+{4W_6\ov (4\p)^2\e}\biggl({\s^2\ov 4\p\m^2}\biggr)^{\!\!\!{-\e/2}}               
-{4S_3\ov (4\p)^4\e^2}\biggl({\s^2\ov 4\p\m^2}\biggr)^{\!\!\!{-\e}}          
\;,           
\eea          
where $W_n$'s ($n=1,4,6$) are the two-loop integrals given in           
Eq.~(\ref{uuu}) and $S_n$'s ($n=1,3$) are the one-loop integrals in           
Eq.~(\ref{www}).           
Likewise, let us define ${\cal{Q}}_n$ as          
\bea          
{\cal{Q}}_n&\equiv&                 
\int_{kpq}{1\ov (k^2+\s^2)^n(p^2+\s^2)[(p+k)^2+\s^2]          
(q^2+\s^2/3)[(q+k)^2+\s^2/3]}\;,\nn          
\eea          
and use the following separation of the pole term      
\bea          
\int_p{1\ov (p^2+\s^2/3)[(p+k)^2+\s^2/3]}               
&=&{1\ov (4\p)^2}\biggl({\s^2/3\ov 4\p\m^2}\biggr)^{\!\!\!{-\e/2}}               
\biggl[{2\ov \e}+G(k^2)\biggr]\;,\label{dcg}          
\eea          
together with Eq.~(\ref{dcf}). Then three-loop integrals $J(b)$ in 
Fig.~1, $K(b)$ in Fig.~2, and $L(b)$ in Fig.~3 are given as          
\bea          
J(b)\equiv{\cal{Q}}_0&=&                
{1\ov (4\p)^4}\biggl({\s^2\ov 4\p\m^2}\biggr)^{\!\!\!{-\e/2}}               
\biggl({\s^2/3\ov 4\p\m^2}\biggr)^{\!\!\!{-\e/2}}               
\int_k F(k^2)G(k^2)\nn\\               
&+&{2W_2\ov (4\p)^2\e}\biggl({\s^2\ov 4\p\m^2}\biggr)^{\!\!\!{-\e/2}}           
+{2W_1\ov (4\p)^2\e}\biggl({\s^2/3\ov 4\p\m^2}\biggr)^{\!\!\!{-\e/2}}          
\;,\nn\\              
K(b)\equiv{\cal{Q}}_1&=&                     
{1\ov (4\p)^4}\biggl({\s^2\ov 4\p\m^2}\biggr)^{\!\!\!{-\e/2}}               
\biggl({\s^2/3\ov 4\p\m^2}\biggr)^{\!\!\!{-\e/2}}               
\int_k{F(k^2)G(k^2)\ov k^2+\s^2}\nn\\               
&+&{2W_5\ov (4\p)^2\e}\biggl({\s^2\ov 4\p\m^2}\biggr)^{\!\!\!{-\e/2}}                
+{2W_4\ov (4\p)^2\e}\biggl({\s^2/3\ov 4\p\m^2}\biggr)^{\!\!\!{-\e/2}}              
-{4S_1\ov (4\p)^4\e^2}\biggl({\s^2\ov 4\p\m^2}\biggr)^{\!\!\!{-\e/2}}               
\biggl({\s^2/3\ov 4\p\m^2}\biggr)^{\!\!\!{-\e/2}}\;,\nn\\            
L(b)\equiv{\cal{Q}}_2&=&                    
{1\ov (4\p)^4}\biggl({\s^2\ov 4\p\m^2}\biggr)^{\!\!\!{-\e/2}}               
\biggl({\s^2/3\ov 4\p\m^2}\biggr)^{\!\!\!{-\e/2}}               
\int_k{F(k^2)G(k^2)\ov (k^2+\s^2)^2}\nn\\               
&+&{2W_8\ov (4\p)^2\e}\biggl({\s^2\ov 4\p\m^2}\biggr)^{\!\!\!{-\e/2}}                
+{2W_6\ov (4\p)^2\e}\biggl({\s^2/3\ov 4\p\m^2}           
\biggr)^{\!\!\!{-\e/2}}               
-{4S_3\ov (4\p)^4\e^2}\biggl({\s^2\ov 4\p\m^2}\biggr)^{\!\!\!{-\e/2}}               
\biggl({\s^2/3\ov 4\p\m^2}\biggr)^{\!\!\!{-\e/2}}\;,\nn\\           
\eea           
where $W_n$'s ($n=1,2,4,5,6,8$) are the two-loop integrals given in           
Eq.~(\ref{uuu}) and $S_n$'s ($n=1,3$) are the one-loop integrals in           
Eq.~(\ref{www}).           
Similarly, defining ${\cal{R}}_n$ as          
\bea          
{\cal{R}}_n&\equiv&                 
\int_{kpq}{1\ov (k^2+\s^2/3)^n(p^2+\s^2)[(p+k)^2+\s^2/3]          
(q^2+\s^2)[(q+k)^2+\s^2/3]}\;,\nn          
\eea          
and using the following separation of the pole part          
\bea          
\int_p{1\ov (p^2+\s^2)[(p+k)^2+\s^2/3]}               
&=&{1\ov (4\p)^2}\biggl({\s^2\ov 4\p\m^2}\biggr)^{\!\!\!{-\e/2}}               
\biggl[{2\ov \e}+H(k^2)\biggr]\;,\label{dce}          
\eea          
we see that three-loop integrals $J(b)$ in Fig.~1, $K(c)$ in Fig.~2, 
and $L(c)$ in Fig.~3 are given as          
\bea          
J(b)&\equiv&{\cal{R}}_0=          
{1\ov (4\p)^4}\biggl({\s^2\ov 4\p\m^2}\biggr)^{\!\!\!{-\e}}             
\int_k \Bigl(H(k^2)\Bigr)^{\!2}+{4W_3\ov (4\p)^2\e}             
\biggl({\s^2\ov 4\p\m^2}\biggr)^{\!\!\!{-\e/2}}\;,\nn\\           
K(c)&\equiv&{\cal{R}}_1=                
{1\ov (4\p)^4}\biggl({\s^2\ov 4\p\m^2}\biggr)^{\!\!\!{-\e}}               
\int_k{[H(k^2)]^2\ov k^2+\s^2/3}               
+{4W_5\ov (4\p)^2\e}\biggl({\s^2\ov 4\p\m^2}\biggr)^{\!\!\!{-\e/2}}               
-{4S_2\ov (4\p)^4\e^2}\biggl({\s^2\ov 4\p\m^2}\biggr)^{\!\!\!{-\e}}\;,\nn\\            
L(c)&\equiv&{\cal{R}}_2=          
{1\ov (4\p)^4}\biggl({\s^2\ov 4\p\m^2}\biggr)^{\!\!\!{-\e}}               
\int_k{[H(k^2)]^2\ov (k^2+\s^2/3)^2}               
+{4W_7\ov (4\p)^2\e}\biggl({\s^2\ov 4\p\m^2}\biggr)^{\!\!\!{-\e/2}}               
-{4S_4\ov (4\p)^4\e^2}\biggl({\s^2\ov 4\p\m^2}           
\biggr)^{\!\!\!{-\e}}\;,          
\eea           
where $W_n$'s ($n=3,5,7$) are the two-loop integrals given in           
Eq.~(\ref{uuu}) and $S_n$'s ($n=2,4$) are the one-loop integrals in           
Eq.~(\ref{www}).          
Finally, we define ${\cal{T}}_n$ as          
\bea          
{\cal{T}}_n&\equiv&                 
\int_{kpq}{1\ov (k^2+\s^2)^n(p^2+\s^2/3)[(p+k)^2+\s^2/3]          
(q^2+\s^2/3)[(q+k)^2+\s^2/3]}\;.\nn          
\eea          
With the decomposition, Eq.~(\ref{dcg}), we obtain the following decomposed           
expressions for three-loop integrals $J(c)$ in Fig.~1, $K(d)$ in Fig.~2, 
and $L(d)$ in Fig.~3:          
\bea          
J(c)&\equiv&{\cal{T}}_0=          
{1\ov (4\p)^4}\biggl({\s^2/3\ov 4\p\m^2}\biggr)^{\!\!\!{-\e}}               
\int_k \Bigl(G(k^2)\Bigr)^2               
+{4W_2\ov (4\p)^2\e}\biggl({\s^2/3\ov 4\p\m^2}           
\biggr)^{\!\!\!{-\e/2}}\;,\nn\\            
K(d)&\equiv&{\cal{T}}_1=          
{1\ov (4\p)^4}\biggl({\s^2/3\ov 4\p\m^2}\biggr)^{\!\!\!{-\e}}               
\int_k{[G(k^2)]^2\ov k^2+\s^2}               
+{4W_5\ov (4\p)^2\e}\biggl({\s^2/3\ov 4\p\m^2}\biggr)^{\!\!\!{-\e/2}}               
-{4S_1\ov (4\p)^4\e^2}\biggl({\s^2/3\ov 4\p\m^2}           
\biggr)^{\!\!\!{-\e}}\;,\nn\\          
L(d)&\equiv&{\cal{T}}_2=          
{1\ov (4\p)^4}\biggl({\s^2/3\ov 4\p\m^2}\biggr)^{\!\!\!{-\e}}               
\int_k{[G(k^2)]^2\ov (k^2+\s^2)^2}               
+{4W_8\ov (4\p)^2\e}\biggl({\s^2/3\ov 4\p\m^2}\biggr)^{\!\!\!{-\e/2}}               
-{4S_3\ov (4\p)^4\e^2}\biggl({\s^2/3\ov 4\p\m^2}           
\biggr)^{\!\!\!{-\e}}\;,          
\eea          
where $W_n$'s ($n=2,5,8$) are the two-loop integrals given in           
Eq.~(\ref{uuu}) and $S_n$'s ($n=1,3$) are the one-loop integrals in           
Eq.~(\ref{www}).

Three integrals in Fig.~4 are relatively simple, though their three          
integration variables ($k$, $p$, and $q$) are overlapped 
in a most complicated way. We investigate first the three-loop integral 
$M(a)$ in Fig.~4:          
\bea           
M(a)&\equiv&\int_{kpq}{1\ov (k^2+\s^2)(p^2+\s^2)(q^2+\s^2)                 
[(k-p)^2+\s^2][(p-q)^2+\s^2][(q-k)^2+\s^2]} \nn\\          
&=&\int_k{I(k^2)\ov k^2+\s^2}\;,\nn          
\eea          
where          
\bea           
I(k^2)=\int_{pq}{1\ov (p^2+\s^2)(q^2+\s^2)[(k-p)^2+\s^2][(q-k)^2+\s^2]          
[(p-q)^2+\s^2]}\;.\nn           
\eea          
The calculation of $M(a)$ can be found in the literature \cite{cmm,pv}.          
The two-loop integration $I(k^2)$ above is finite because it           
has negative degree of divergence and no subdivergences. The divergences           
of $M(a)$ can only arise from integration over large $k^2$. Thus it is          
sufficient to know the behavior of $I(k^2)$ at large $k^2$. Details can be          
found in Ref.~\cite{cmm}. In the Appendix B of           
Ref.~\cite{pv} the method of Kotikov \cite{kk} is used for the integration      
of $M(a)$.          
The result is           
\bea          
M(a)&=&{1\ov (4\p)^6}\biggl({\s^2\ov 4\p\m^2}\biggr)^{\!\!\!-3\e/2}                 
\biggl[{4\z(3)\ov \e}\biggr]\;.\nn          
\eea          
Other two three-loop integrals $M(b)$ and $M(c)$ in Fig.~4 are equal           
to $M(a)$ within the desired order. This can be seen easily by expanding          
the propagators with one third of $\s^2$ in the following          
fashion:          
\beas          
{1\ov p^2+\s^2/3}={1\ov p^2+\s^2-(2/3)\s^2}={1\ov p^2+\s^2}\sum_{\ell=0}^\8          
\biggl({2\s^2\ov 3}\biggr)^{\!\!\ell}\biggl({1\ov p^2+\s^2}\biggr)^{\!\!\ell}\;.          
\eeas          
Consequently we have           
\bea          
M(b)&\equiv&               
\int_{kpq}{1\ov (k^2+\s^2/3)(p^2+\s^2/3)(q^2+\s^2/3)                 
[(k-p)^2+\s^2][(p-q)^2+\s^2][(q-k)^2+\s^2]}\nn\\                
&=&M(a)+O(\e^0)\;,\nn\\                  
M(c)&\equiv&               
\int_{kpq}{1\ov (k^2+\s^2)(p^2+\s^2/3)(q^2+\s^2/3)                 
[(k-p)^2+\s^2/3][(p-q)^2+\s^2][(q-k)^2+\s^2/3]}\nn\\                
&=&M(a)+O(\e^0)\;. \nn             
\eea

\section{Integrals Containing $F$, $G$, and $H$}                 
In order to calculate the $k$ integrals containing $F(k^2)$, $G(k^2)$, and       
$H(k^2)$, let us first investigate the following decomposition of       
a propagator-type one-loop integral:        
\bea          
\int_p{1\ov (p^2+\s^2)^{k_{1}}[(p+k)^2+\s^2]^{k_{2}}}&=&          
{(\s^2)^{4-2(k_1+k_2)}\ov (4\p)^2}\biggl({\s^2\ov 4\p\m^2}\biggr)^{\!\!\!{
-\e/2}}               
\biggl[{2\ov \e}\d_{1k_{1}}\d_{1k_{2}}     
+F^{\rm ren}_{k_{1}k_{2}}(k^2)\biggr]\;.          
\label{k1k2}          
\eea          
The asymptotic behavior of $F^{\rm ren}_{k_{1}k_{2}}(k^2)$ is given      
as \cite{cmm,bd}         
\bea          
F^{\rm ren}_{k_{1}k_{2}}(k^2)=     
\sum_{r=0}^\8\biggl[\biggl({\s^2\ov k^2}\biggr)^{\!\!r}          
a^{\rm ren}(k_1,k_2,r)          
+\biggl({\s^2\ov k^2}\biggr)^{\!\!r+\e/2}b(k_1,k_2,r)\biggr]\;,         
\label{asf}         
\eea          
where          
\beas          
a^{\rm ren}(k_1,k_2,r) &=& \left\{ \begin{array}{ll}          
\vspace{0.3cm}\displaystyle{-{2\ov \e}\d_{1k_{1}}\d_{1k_{2}}     
\d_{0r}}~~~\Bigl(\mbox{when~~$r<k_1$~~~          
or~~${k_1+k_2\ov 2}\le r<k_2$}\Bigr)\;,\nn\\          
\displaystyle{-{2\ov \e}\d_{1k_{1}}\d_{1k_{2}}\d_{0r}          
+{(-1)^{r-k_1}(r-1)!(k_2-r-1)! \G(k_1+k_2-r-2+\e/2)\ov          
(k_1-1)!(k_2-1)!(r-k_1)!(k_1+k_2-2r-1)!}}\nn\\          
\vspace{0.3cm}~~~~~~~~~~~~~~~~~~~~~~~~~~~~~~~~~~~~~~~~~~~~~~~~~~~~          
\Bigl({\rm when~~}k_1\le r<{k_1+k_2\ov 2}\Bigr)\;,\nn\\          
\displaystyle{-{2\ov \e}\d_{1k_{1}}\d_{1k_{2}}\d_{0r}     
+\fr{2 (r-1)!(2r-k_1-k_2)!           
\G(k_1+k_2-r-2+\e/2)}{(k_1-1)!(k_2-1)!(r-k_1)!(r-k_2)!}}\nn\\          
~~~~~~~~~~~~~~~~~~~~~~~~~~~~~~~~~~~~~~~~~~~~~~~~~~~~~~~~~~~~~~          
\Bigl({\rm when~~}r\ge k_2\Bigr)\;,          
\end{array} \right.\nn\\           
b(k_1,k_2,r)&=&\left\{ \begin{array}{ll}          
\vspace{0.3cm}~~0~~~~~~~          
\Bigl({\rm when~~}r<k_1+k_2-2\Bigr)\;,\nn\\          
\displaystyle{          
{ (-1)^{r-k_1-k_2+2} \G(k_1-r-\e/2)\G(k_2-r-\e/2)\G(r+\e/2)\ov          
(k_1-1)!(k_2-1)!(r-k_1-k_2+2)!\G(k_1+k_2-2r-\e)}}\nn\\          
~~~~~~~~~~~~~~~~~~~~~~~~~~~~~~~~~~~~~~~~~~~~~~~~~~~          
\Bigl({\rm when~~}r\ge k_1+k_2-2\Bigr)\;.          
\end{array} \right.           
\eeas          
Note that $a^{\rm ren}(1,1,0)$ does contain a simple pole which cancels      
the pole of $b(1,1,0)$ in the expression $F^{\rm ren}_{k_{1}k_{2}}(k^2)$.      
Thus we see that the asymptotic behavior of $F(k^2)$ in Eq.~(\ref{dcf}) is       
given as          
\bea          
F(k^2)=\sum_{r=0}^\8\biggl[\biggl({\s^2\ov k^2}\biggr)^{\!\!r}a^{\rm ren}(1,1,r)          
+\biggl({\s^2\ov k^2}\biggr)^{\!\!r+\e/2}b(1,1,r)\biggr]\;.\label{fays}          
\eea          
Using this equation we obtain          
\bea          
{1\ov (4\p)^4}\biggl({\s^2\ov 4\p\m^2}\biggr)^{\!\!\!{-\e}}               
\int_k{[F(k^2)]^2\ov (k^2+\s^2)^n}&=&{(\s^2)^{2-n}\ov (4\p)^6}          
\biggl({\s^2\ov 4\p\m^2}\biggr)^{\!\!\!{-3\e/2}}{2\ov\e\G(2-\e/2)}\nn\\          
&&\!\!\!\!\!\!\!\!\!\!\!\!\!\!\!\!\!\!\!\!\!\!\!\!\!\!\!\!\!\!\!\!\!\!\!\!\!          
\!\!\!\!\!\!\!\!\!\!\!\!\!\!\!\!\!\!\!\!          
\times\sum_{r_{1}=0}^{2-n}\,\,\,\sum_{r_{2}=0}^{2-n-r_{1}}\biggl(          
{-n\atop 2-n-r_1-r_2}\biggr)\biggl[a^{\rm ren}(1,1,r_1)a^{\rm ren}(1,1,r_2)          
+{1\ov 3}b(1,1,r_1)b(1,1,r_2)\nn\\          
&&\!\!\!\!\!\!\!\!\!\!\!\!\!\!\!\!\!\!\!\!\!\!\!\!\!\!\!          
\!\!\!\!\!\!\!\!\!\!\!\!\!\!\!\!\!\!\!\!          
+{1\ov 2}\Bigl\{a^{\rm ren}(1,1,r_1)b(1,1,r_2)+          
b(1,1,r_1)a^{\rm ren}(1,1,r_2)\Bigr\}\biggr]+\mbox{[finite terms]}\;,          
\label{ff}          
\eea          
where the symbol $({a\atop b})$ is the expansion coefficient of           
$(1+x)^a=\sum_{b=0}^\8({a\atop b})x^b$. Details of the step leading to           
Eq.~(\ref{ff}) can be found in Ref.~\cite{cmm}. Therefore, the calculations        
of three integrals in ${\cal{P}}_0$, ${\cal{P}}_1$, and ${\cal{P}}_2$ can      
be regarded as           
simple quotations of Ref.~\cite{cmm}:             
\bea          
&&{1\ov (4\p)^4}\biggl({\s^2\ov 4\p\m^2}\biggr)^{\!\!\!{-\e}}               
\int_k \Bigl(F(k^2)\Bigr)^{\!2}=\Omega_2\biggl[{1\ov \e^2}               
\biggl\{{44\ov 3}-8\g\biggr\}+{1\ov \e}\biggl\{23-30\g+10\g^2+{\p^2\ov 3}               
\biggr\}+O(\e^0)\biggr]\;,\nn\\           
&&{1\ov (4\p)^4}\biggl({\s^2\ov 4\p\m^2}\biggr)^{\!\!\!{-\e}}               
\int_k{[F(k^2)]^2\ov k^2+\s^2}=\Omega_1\biggl[{8\ov \e^3}+{1\ov \e^2}               
\biggl\{{28\ov 3}-8\g\biggr\}+{1\ov \e}\biggl\{-{14\ov 3}               
+2\g^2+{\p^2\ov 3}\biggr\}+O(\e^0)\biggr]\;,\nn\\           
&&{1\ov (4\p)^4}\biggl({\s^2\ov 4\p\m^2}\biggr)^{\!\!\!{-\e}}               
\int_k{[F(k^2)]^2\ov (k^2+\s^2)^2}=\Omega_0\biggl[{8\ov 3\e^3}               
-{4\ov 3\e^2}-{2\ov 3\e}+O(\e^0)\biggr]\;.           
\eea          
In the above equation the overall multiplying factors are defined as              
\beas               
\Omega_0={1\ov (4\p)^6}\biggl({\s^2\ov 4\p\m^2}           
\biggr)^{\!\!\!{-3\e/2}}\;,~~~               
\Omega_1={\s^2\ov (4\p)^6}\biggl({\s^2\ov 4\p\m^2}           
\biggr)^{\!\!\!{-3\e/2}}\;,             
~~~               
\Omega_2={\s^4\ov (4\p)^6}\biggl({\s^2\ov 4\p\m^2}\biggr)^{\!\!\!{-3\e/2}}\;,              
\eeas              
and $\g$ is the usual Euler constant, $\g=0.5772156649\cdots$.

Likewise, the asymptotic behavior of $G(k^2)$ in Eq.~(\ref{dcg}) is given as          
\bea          
G(k^2)=\sum_{r=0}^\8\biggl[\biggl({\s^2\ov k^2}\biggr)^{\!\!r}          
a^{\rm ren}(1,1,r)\biggl({1\ov 3}\biggr)^{\!\!r}          
+\biggl({\s^2\ov k^2}\biggr)^{\!\!r+\e/2}b(1,1,r)          
\biggl({1\ov 3}\biggr)^{\!\!r+\e/2}\biggr]\;. \label{gays}         
\eea          
Using this equation together with Eq.~(\ref{fays}), we obtain            
\bea          
&&{1\ov (4\p)^4}\biggl({\s^2\ov 4\p\m^2}\biggr)^{\!\!\!{-\e/2}}               
\biggl({\s^2/3\ov 4\p\m^2}\biggr)^{\!\!\!{-\e/2}}               
\int_k{F(k^2)G(k^2)\ov (k^2+\s^2)^n}={(\s^2)^{2-n}\ov (4\p)^6}          
\biggl({\s^2\ov 4\p\m^2}\biggr)^{\!\!\!-\e}          
\biggl({\s^2/3\ov 4\p\m^2}\biggr)^{\!\!\!-\e/2}          
{2\ov\e\G(2-\e/2)}\nn\\          
&&~~~~~~~~~~~~~          
\times\sum_{r_{1}=0}^{2-n}\,\,\,\sum_{r_{2}=0}^{2-n-r_{1}}\biggl(          
{-n\atop 2-n-r_1-r_2}\biggr)\biggl[a^{\rm ren}(1,1,r_1)a^{\rm ren}(1,1,r_2)          
\biggl({1\ov 3}\biggr)^{\!\!r_{2}}\nn\\          
&&~~~~~~~~~~~~~~~~~~~          
+{1\ov 3}b(1,1,r_1)b(1,1,r_2)          
\biggl({1\ov 3}\biggr)^{\!\!r_{2}+\e/2}\nn\\          
&&~~~~~~~~~~~~~~~~~~~~~~~          
+{1\ov 2}\biggl\{a^{\rm ren}(1,1,r_1)b(1,1,r_2)          
\biggl({1\ov 3}\biggr)^{\!\!r_{2}+\e/2}+          
b(1,1,r_1)a^{\rm ren}(1,1,r_2)\biggl({1\ov 3}\biggr)^{\!\!r_{2}}\biggr\}          
\biggr]\nn\\          
&&~~~~~~~~~~~~~~~~~~~~~~~~~~~~~~          
+\mbox{[finite terms]}\;.          
\eea          
From this equation three integrals corresponding to $n=0,1,2$ are           
evaluated as follows:          
\bea          
&&{1\ov (4\p)^4}\biggl({\s^2\ov 4\p\m^2}\biggr)^{\!\!\!{-\e/2}}               
\biggl({\s^2/3\ov 4\p\m^2}\biggr)^{\!\!\!{-\e/2}}               
\int_k F(k^2)G(k^2)=\Omega_2\biggl[-{64\ov 27\e^3}+{1\ov \e^2}               
\biggl\{{92\ov 27}-{8\ov 9}\g-{16\ov 9}\ln 3\biggr\}\nn\\               
&&~~~~~~~~~~~~~~~~~~~~~~               
+{1\ov \e}\biggl\{{185\ov 27}-{86\ov 9}\g+{26\ov 9}\g^2+{\p^2\ov 27}               
+{11\ov 9}\ln 3+{2\ov 9}\g\ln 3-{4\ov 9}\ln^2 3\biggr\}+O(\e^0)           
\biggr]\;,\nn\\            
&&{1\ov (4\p)^4}\biggl({\s^2\ov 4\p\m^2}\biggr)^{\!\!\!{-\e/2}}               
\biggl({\s^2/3\ov 4\p\m^2}\biggr)^{\!\!\!{-\e/2}}               
\int_k{F(k^2)G(k^2)\ov k^2+\s^2}=\Omega_1\biggl[{40\ov 9\e^3}               
+{1\ov \e^2}\biggl\{{20\ov 3}-{16\ov 3}\g+{10\ov 3}\ln 3\biggr\}\nn\\               
&&~~~~~~~~~~~~~~~~~~~~~~~~~~~             
+{1\ov \e}\biggl\{-{26\ov 9}+{4\ov 3}\g^2+{2\ov 9}\p^2               
+7\ln 3-{14\ov 3}\g\ln 3+{5\ov 6}\ln^2 3\biggr\}+O(\e^0)\biggr]\;,\nn\\           
&&{1\ov (4\p)^4}\biggl({\s^2\ov 4\p\m^2}\biggr)^{\!\!\!{-\e/2}}               
\biggl({\s^2/3\ov 4\p\m^2}\biggr)^{\!\!\!{-\e/2}}               
\int_k{F(k^2)G(k^2)\ov (k^2+\s^2)^2}=\Omega_0\biggl[{8\ov 3\e^3}               
+{1\ov \e^2}\biggl\{-{4\ov 3}+2\ln 3\biggr\}            
\nn\\   &&~~~~~~~~~~~~~~~~~~~~~~~~~~~~~~~~~~~~~~~~~~~~~~~~~~~~~~~~~~~~~~~~               
+{1\ov \e}\biggl\{-{2\ov 3}-\ln 3+{1\ov 2}\ln^2 3\biggr\}               
+O(\e^0)\biggr]\;.          
\eea

Similarly the asymptotic behavior of $H(k^2)$ is given as          
\bea          
H(k^2)=\sum_{\a=0}^\8\sum_{r=0}^\8          
\biggl({2\s^2\ov 3}\biggr)^{\!\!\a}          
\biggl[\biggl({\s^2\ov k^2}\biggr)^{\!\!r}a^{\rm ren}(1,1+\a,r)          
+\biggl({\s^2\ov k^2}\biggr)^{\!\!r+\e/2}b(1,1+\a,r)\biggr]\;.\label{h}          
\eea          
This can be shown readily by writing          
\bea          
\int_p{1\ov (p^2+\s^2)[(p+k)^2+\s^2/3]}               
&=&\sum_{\a=0}^\8\biggl({2\s^2\ov 3}\biggr)^{\!\!\a}          
\int_p{1\ov (p^2+\s^2)[(p+k)^2+\s^2]^{1+\a}}\;,\label{ep}         
\eea          
and using Eq.~(\ref{k1k2}). Note that there exists a one-loop integration          
formula for the diagrams with arbitrary two different masses in propagators          
obtained by Boos and Davydychev \cite{bd}. See Eq.~(21) of this reference.         
Applying directly this formula expressed in terms of Appell's hypergeometric          
function $F_4$ of two variables  to our calculation is more difficult than        
using the expansion of Eq.~(\ref{ep}) and the formulas of Eqs.~(\ref{k1k2})        
and (\ref{asf}).         
Using Eq.~(\ref{h}) we arrive at            
\bea          
{1\ov (4\p)^4}\biggl({\s^2\ov 4\p\m^2}\biggr)^{\!\!\!{-\e}}          
\int_k{[H(k^2)]^2\ov (k^2+\s^2/3)^n}&=&{(\s^2)^{2-n}\ov (4\p)^6}          
\biggl({\s^2\ov 4\p\m^2}\biggr)^{\!\!\!{-3\e/2}}{2\ov\e\G(2-\e/2)}\nn\\          
&&\!\!\!\!\!\!\!\!\!\!\!\!\!\!\!\!\!\!\!\!\!\!\!\!\!\!\!\!\!\!\!\!\!\!\!\!\!          
\!\!\!\!\!\!\!\!\!\!\!\!\!\!\!\!\!\!\!\!\!\!\!\!\!\!\!\!\!\!\!\!          
\times\sum_{\a=0}^\8\sum_{\b=0}^\8\biggl({2\ov 3}\biggr)^{\a+\b}\,\,          
\sum_{r_{1}=0}^{2-n}\,\,\,\sum_{r_{2}=0}^{2-n-r_{1}}          
\biggl({1\ov 3}\biggr)^{2-n-r_{1}-r_{2}}\biggl(          
{-n\atop 2-n-r_1-r_2}\biggr)\nn\\          
&&\!\!\!\!\!\!\!\!\!\!\!\!\!\!\!\!\!\!\!\!\!\!\!\!\!\!\!\!\!\!\!\!\!\!\!\!\!          
\!\!\!\!\!\!\!\!\!\!\!\!\!\!\!\!\!\!\!\!\!\!\!          
\times\biggl[a^{\rm ren}(1,1+\a,r_1)          
a^{\rm ren}(1,1+\b,r_2)+{1\ov 3}b(1,1+\a,r_1)b(1,1+\b,r_2)\nn\\          
&&\!\!\!\!\!\!\!\!\!\!\!\!\!\!\!\!\!\!\!\!\!\!\!\!\!\!\!          
\!\!\!\!\!\!\!\!\!\!\!\!\!\!\!\!\!\!\!\!\!\!\!\!          
+{1\ov 2}\Bigl\{a^{\rm ren}(1,1+\a,r_1)b(1,1+\b,r_2)+          
b(1,1+\a,r_1)a^{\rm ren}(1,1+\b,r_2)\Bigr\}\biggr]\nn\\          
&&\!\!\!\!\!\!\!\!\!\!\!\!\!\!\!\!\!\!\!\!\!\!\!\!\!\!\!          
\!\!\!\!\!\!\!\!\!\!\!\!\!\!         
+\mbox{[finite terms]}\;, \label{hh}         
\eea               
from which three integrals for $n=0,1,2$ are calculated as follows:          
\bea             
&&{1\ov (4\p)^4}\biggl({\s^2\ov 4\p\m^2}\biggr)^{\!\!\!{-\e}}             
\int_k \Bigl(H(k^2)\Bigr)^{\!2}=\Omega_2\biggl[{32\ov 27\e^3}+{1\ov \e^2}               
\biggl\{{188\ov 27}-{40\ov 9}\g+{4\ov 9}\ln 3\biggr\}\nn\\               
&&~~~~~~~~~~~~~~~~~~              
+{1\ov \e}\biggl\{{257\ov 27}-{118\ov 9}\g+{14\ov 3}\g^2+{5\ov 27}\p^2               
+{31\ov 9}\ln 3-2\g\ln 3+{1\ov 3}\ln^2 3\biggr\}+O(\e^0)\biggr]\;,\nn\\          
&&{1\ov (4\p)^4}\biggl({\s^2\ov 4\p\m^2}\biggr)^{\!\!\!{-\e}}               
\int_k{[H(k^2)]^2\ov k^2+\s^2/3}=\Omega_1\biggl[{56\ov 9\e^3}               
+{1\ov \e^2}\biggl\{{52\ov 9}-{16\ov 3}\g+{4\ov 3}\ln 3\biggr\}\nn\\               
&&~~~~~~~~~~~~~~~~~~~~~~~~~              
+{1\ov \e}\biggl\{-{10\ov 3}+{4\ov 3}\g^2               
+{2\ov 9}\p^2-{2\ov 3}\g\ln 3    
+{1\ov 3}\ln^2 3\biggr\}+O(\e^0)\biggr]\;,\nn\\            
&&{1\ov (4\p)^4}\biggl({\s^2\ov 4\p\m^2}\biggr)^{\!\!\!{-\e}}               
\int_k{[H(k^2)]^2\ov (k^2+\s^2/3)^2}=\Omega_0\biggl[{8\ov 3\e^3}               
-{4\ov 3\e^2}-{2\ov 3\e}+O(\e^0)\biggr]\;.\nn          
\eea         
In dealing with the infinite sums in Eq.~(\ref{hh}) $(n=0,1$)          
we have used the following relations:         
\bea         
\sum_{\a=1}^\8\biggl({2\ov 3}\biggr)^{\!\!\a}{\y(\a)\ov \a(\a+1)}=         
-{1\ov 2}\sum_{\a=1}^\8\biggl({2\ov 3}\biggr)^{\!\!\a}\,{\y(\a)\ov \a}         
+1-\g-{1\ov 2}\ln 3\;,\nn\\         
\sum_{\a=1}^\8\biggl({2\ov 3}\biggr)^{\!\!\a}{\y(\a)\ov \a(\a+1)(\a+2)}=         
{1\ov 8}\sum_{\a=1}^\8\biggl({2\ov 3}\biggr)^{\!\!\a}\,{\y(\a)\ov \a}         
-{1\ov 4}+{3\ov 16}\ln 3\;,\nn         
\eea         
with          
\bea         
\sum_{\a=1}^\8\biggl({2\ov 3}\biggr)^{\!\!\a}\,{\y(\a)\ov \a}=         
-\g\ln 3+{1\ov 2}\ln^2 3\;.\nn         
\eea         
         
Finally, using the asymptotic form of $G(k^2)$, Eq.~(\ref{gays}), we obtain         
\bea          
&&{1\ov (4\p)^4}\biggl({\s^2/3\ov 4\p\m^2}\biggr)^{\!\!\!{-\e}}          
\int_k{[G(k^2)]^2\ov (k^2+\s^2)^2}={(\s^2)^{2-n}\ov (4\p)^6}          
\biggl({\s^2\ov 4\p\m^2}\biggr)^{\!\!\!-\e/2}          
\biggl({\s^2/3\ov 4\p\m^2}\biggr)^{\!\!\!-\e}          
{2\ov\e\G(2-\e/2)}\nn\\          
&&~~~~~~~~~~~~~          
\times\sum_{r_{1}=0}^{2-n}\,\,\,\sum_{r_{2}=0}^{2-n-r_{1}}\biggl(          
{-n\atop 2-n-r_1-r_2}\biggr)\biggl[a^{\rm ren}(1,1,r_1)a^{\rm ren}(1,1,r_2)          
\biggl({1\ov 3}\biggr)^{\!\!r_{1}+r_{2}}\nn\\          
&&~~~~~~~~~~~~~~~~~~~          
+{1\ov 3}b(1,1,r_1)b(1,1,r_2)          
\biggl({1\ov 3}\biggr)^{\!\!r_{1}+r_{2}+\e}\nn\\          
&&~~~~~~~~~~~~~~~~~~~~~~~          
+{1\ov 2}\biggl\{a^{\rm ren}(1,1,r_1)b(1,1,r_2)          
+b(1,1,r_1)a^{\rm ren}(1,1,r_2)\biggr\}          
\biggl({1\ov 3}\biggr)^{\!\!r_{1}+r_{2}+\e/2}\biggr]\nn\\          
&&~~~~~~~~~~~~~~~~~~~~~~~~~~~~~~          
+\mbox{[finite terms]}\;,\nn          
\eea          
and after substitutions of the appropriate values          
for $a^{\rm ren}(1,1,r)$ and $b(1,1,r)$ we end up with         
\bea             
&&{1\ov (4\p)^4}\biggl({\s^2/3\ov 4\p\m^2}\biggr)^{\!\!\!{-\e}}               
\int_k \Bigl(G(k^2)\Bigr)^{\!2}=\Omega_2\biggl[{1\ov \e^2}               
\biggl\{{44\ov 27}-{8\ov 9}\g\biggr\}\nn\\               
&&~~~~~~~~~~~~~~~~~~~~~~~~~~~~~~~+{1\ov \e}\biggl\{{23\ov 9}               
-{10\ov 3}\g+{10\ov 9}\g^2+{\p^2\ov 27}+{22\ov 9}\ln 3-{4\ov 3}\g\ln 3               
\biggr\}+O(\e^0)\biggr]\;,\nn\\               
&&{1\ov (4\p)^4}\biggl({\s^2/3\ov 4\p\m^2}\biggr)^{\!\!\!{-\e}}               
\int_k{[G(k^2)]^2\ov k^2+\s^2}=\Omega_1\biggl[{8\ov 9\e^3}               
+{1\ov \e^2}\biggl\{4-{8\ov 3}\g+{4\ov 3}\ln 3\biggr\}\nn\\               
&&~~~~~~~~~~~~~~~~~~~~~~~~~~~~               
+{1\ov \e}\biggl\{-{10\ov 9}+{2\ov 3}\g^2+{\p^2\ov 9}               
+6\ln 3-4\g\ln 3+\ln^2 3\biggr\}+O(\e^0)\biggr]\;,\nn\\          
&&{1\ov (4\p)^4}\biggl({\s^2/3\ov 4\p\m^2}\biggr)^{\!\!\!{-\e}}          
\int_k{[G(k^2)]^2\ov (k^2+\s^2)^2}=\Omega_0\biggl[{8\ov 3\e^3}               
+{1\ov \e^2}\biggl\{-{4\ov 3}+4\ln 3\biggr\}\nn\\            
&&~~~~~~~~~~~~~~~~~~~~~~~~~~~~~~~~~~~~~~~~~~~~~~~~~~~              
+{1\ov \e}\biggl\{-{2\ov 3}-2\ln 3+3\ln^2 3\biggr\}+O(\e^0)\biggr]\;.\nn               
\eea

\section{Summary}                 
We summarize all results calculated in the previous sections           
by listing them in the $\e$-expanded forms, but leaving the $\Omega_{0,1,2}$           
factors unexpanded. Our desired accuracy of three-loop integrals is to      
calculate up to the order of $\e^{-1}$ apart from an overall       
multiplying factor $(\s^2/(4\p\m^2))^{-3\e/2}$. When this overall      
multiplying factor is expanded in powers of $\e$, the resulting     
accuracy is up to finite terms containing the logarithm of mass squared,       
$\ln (\s^2/(4\p\m^2))$. The list reads as follows:               
\bea             
J(a)&=&\Omega_2\biggl[{16\ov \e^3}+{1\ov \e^2}\biggl\{{92\ov 3}           
-24\g\biggr\}                 
+{1\ov \e}\biggl\{35-46\g+18\g^2+\p^2\biggr\}\biggr]\;,\nn\\            
J(b)&=&\Omega_2               
\biggl[{176\ov 27\e^3}+{1\ov \e^2}\biggl\{{332\ov 27}-{88\ov 9}\g+               
{28\ov 9}\ln3\biggr\}            
\nn\\    &&~~~~~~            
+{1\ov \e}\biggl\{{365\ov 27}-{166\ov 9}\g+{22\ov 3}\g^2               
+{11\ov 27}\p^2+{55\ov 9}\ln 3-{14\ov 3}\g\ln 3+\ln^2 3\biggr\}           
\biggr]\;,\nn\\                 
J(c)&=&\Omega_2               
\biggl[{16\ov 9\e^3}+{1\ov \e^2}\biggl\{{92\ov 27}-{8\ov 3}\g+{8\ov 3}                 
\ln 3\biggr\}            
\nn\\   &&~~~~~~            
+{1\ov \e}\biggl\{{35\ov 9}-{46\ov 9}\g+2\g^2+{\p^2\ov 9}                 
+{46\ov 9}\ln 3-4\g\ln 3+2\ln^2 3\biggr\}\biggr]\;,\nn\\                 
K(a)&=&\Omega_1\biggl[-{8\ov \e^3}+{1\ov \e^2}\biggl\{-{68\ov 3}           
+12\g\biggr\}                 
+{1\ov \e}\biggl\{-{134\ov 3}-12A+34\g-9\g^2-{\p^2\ov 2}\biggr\}           
\biggr]\;,\nn\\                 
K(b)&=&\Omega_1\biggl[-{56\ov 9\e^3}+{1\ov \e^2}\biggl\{-{52\ov 3}           
+{28\ov 3}\g               
-{4\ov 3}\ln 3\biggr\}            
\nn\\      &&~~~~~~            
+{1\ov \e}\biggl\{-{302\ov 9}-6A-{2\ov 3}B+26\g-7\g^2-{7\ov 18}\p^2               
-4\ln 3+2\g\ln 3\biggr\}\biggr]\;,\nn\\                 
K(c)&=&\Omega_1               
\biggl[-{40\ov 9\e^3}+{1\ov \e^2}\biggl\{-{116\ov 9}+{20\ov 3}\g           
-{8\ov 3}\ln 3               
\biggr\}\nn\\               
&&~~~~~~+{1\ov \e}\biggl\{-26-{4\ov 3}B+{58\ov 3}\g-5\g^2               
-{5\ov 18}\p^2+4\g\ln 3-{22\ov 3}\ln 3\biggr\}               
\biggr]\;,\nn\\                 
K(d)&=&\Omega_1\biggl[-{40\ov 9\e^3}+{1\ov \e^2}\biggl\{-12+{20\ov 3}\g               
-{8\ov 3}\ln 3\biggr\}            
\nn\\   &&~~~~~~            
+{1\ov \e}\biggl\{-{202\ov 9}-{4\ov 3}B+18\g-5\g^2-{5\ov 18}\p^2-8\ln 3                 
+4\g\ln 3\biggr\}\biggr]\;,\nn\\                 
L(a)&=&\Omega_0\biggl[{8\ov 3\e^3}+{1\ov \e^2}\biggl\{{8\ov 3}-4\g\biggr\}                 
+{1\ov \e}\biggl\{{4\ov 3 }+4A-4\g+3\g^2+{\p^2\ov 6}\biggr\}           
\biggr]\;,\nn\\                 
L(b)&=&\Omega_0\biggl[{8\ov 3\e^3}+{1\ov \e^2}\biggl\{{8\ov 3}-4\g\biggr\}                 
+{1\ov \e}\biggl\{{4\ov 3 }+2A+2C-4\g+3\g^2+{\p^2\ov 6}\biggr\}           
\biggr]\;,\nn\\                 
L(c)&=&\Omega_0               
\biggl[{8\ov 3\e^3}+{1\ov \e^2}\biggl\{{8\ov 3}-4\g+4\ln 3\biggr\}\nn\\            
&&~~~~~~            
+{1\ov \e}\biggl\{{4\ov 3 }+4B-4\g+3\g^2+{\p^2\ov 6}+4\ln 3-6\g\ln 3               
+3\ln^2 3\biggr\}\biggr]\;,\nn\\                 
L(d)&=&\Omega_0\biggl[{8\ov 3\e^3}+{1\ov \e^2}\biggl\{{8\ov 3}-4\g\biggr\}                 
+{1\ov \e}\biggl\{{4\ov 3}+4C-4\g+3\g^2+{\p^2\ov 6}\biggr\}\biggr]\;,\nn\\                 
M(a)&=&M(b)=M(c)=\Omega_0\biggl[{4\z(3)\ov \e}\biggr]\;.  \label{sum}          
\eea                 
In the above equation $A$, $B$, and $C$ are constants whose        
numerical values are given in Eq.~(\ref{abc}).

Now we close this section with three-loop effective potential of 
massless $O(N)$ $\f^4$ theory, calculated using our integration results of three-loop 
diagrams. The renormalization is highly involved, but straightforward \cite{jkps}:
\bea 
V_{\rm eff}(\fh_a)&=&
 \Biggl[{\l\ov 4!}\fh^4\Biggr] 
+{\l^2\fh^4\ov (4\p)^2}\Biggl[-{25\ov 96}+{1\ov 16}   
\ln\biggl({\fh^2\ov M^2}\biggr)   
+\~{N}\biggl\{-{25\ov 864}+{1\ov 144}\ln\biggl({\fh^2\ov M^2}\biggr)   
\biggr\}\Biggr]\nn\\   
&+&{\l^3\fh^4\ov (4\p)^4}\Biggl[{55\ov 24}-{13\ov 16}   
\ln\biggl({\fh^2\ov M^2}\biggr)+{3\ov 32}   
\ln^2\biggl({\fh^2\ov M^2}\biggr)+\~{N}\biggl\{{635\ov 1296}-{19\ov 108}   
\ln\biggl({\fh^2\ov M^2}\biggr)\nn\\ 
&+&{1\ov 48}\ln^2\biggl({\fh^2\ov M^2}\biggr)\biggr\} 
+\~{N}^2\biggl\{{85\ov 3888}-{11\ov 1296}\ln\biggl({\fh^2\ov M^2}\biggr)   
+{1\ov 864}\ln^2\biggl({\fh^2\ov M^2}\biggr)\biggr\}\Biggr]\nn\\
&+&{\l^4\fh^4\ov (4\p)^6}\Biggl[   
-{27035\ov 1152}-{25\ov 32}A-{25\ov 24}\z(3) 
+\biggl({1957\ov 192}+{3\ov 16}A
+{\z(3)\ov 4}\biggr)\ln\biggl({\fh^2\ov M^2}\biggr)\nn\\
&-&{359\ov 192}\ln^2\biggl({\fh^2\ov M^2}\biggr)
+{9\ov 64}\ln^3\biggl({\fh^2\ov M^2}\biggr)
+\~{N}\biggl\{-{228725\ov 31104}-{125\ov 864}A
-{125\ov 648}\z(3)\nn\\
&+&{175\ov 5184}\ln 3+\biggl({16769\ov 5184}
+{5\ov 144}A+{5\ov 108}\z(3)
-{7\ov 864}\ln 3\biggr)\ln\biggl({\fh^2\ov M^2}\biggr)\nn\\
&-&{523\ov 864}\ln^2\biggl({\fh^2\ov M^2}\biggr)
+{3\ov 64}\ln^3\biggl({\fh^2\ov M^2}\biggr)\biggr\}
+\~{N}^2\biggl\{-{62105\ov 93312}-{25\ov 3888}A
+{175\ov 15552}\ln 3\nn\\
&+&\biggl({4775\ov 15552}+{A\ov 648}
-{7\ov 2592}\ln 3\biggr)\ln\biggl({\fh^2\ov M^2}\biggr)
-{319\ov 5184}\ln^2\biggl({\fh^2\ov M^2}\biggr)+{1\ov 192}
\ln^3\biggl({\fh^2\ov M^2}\biggr)\biggr\}\nn\\
&+&\~{N}^3\biggl\{-{4655\ov 279936}
+{395\ov 46656}\ln\biggl({\fh^2\ov M^2}\biggr)
-{5\ov 2592}\ln^2\biggl({\fh^2\ov M^2}\biggr)+{1\ov 5184}   
\ln^3\biggl({\fh^2\ov M^2}\biggr)\biggr\}\Biggr]\;,\label{3p}
\eea  
where $\~N=N-1$.
Up to two-loop level, Eq.~(\ref{3p}) completely coincides with the 
existing calculation in Ref.~\cite{jk}. (Note that in this reference though 
regularization method differs from ours, the same renormalization conditions 
are used.) The above result is exact in parameter $N$ up to three-loop order. 
In order to see that three-loop part of our calculation is not erroneous,
we compare the leading order part of Eq.~(\ref{3p}), when written down in 
$1/N$ expansion formalism with a simple replacement $\l \rar \l/N$, with the 
following extracted overlap part of the large-$N$ limit calculation \cite{cjp}:
\bea
V_{\rm CJP}&=&
{\l\ov 4!N}\f^4+{\l^2\f^4\ov (4\p)^2N}\Biggl[-{1\ov 288}
+{1\ov 144}\ln\biggl({\f^2\ov M^2}\biggr) \Biggr]\nn\\   
&+&{\l^3\f^4\ov (4\p)^4N}\Biggl[   
{1\ov 864}\ln^2\biggl({\f^2\ov M^2}\biggr)\Biggr]
+{\l^4\f^4\ov (4\p)^6 N}\Biggl[   
{1\ov 5184}\ln^2\biggl({\f^2\ov M^2}\biggr)   
+{1\ov 5184}\ln^3\biggl({\f^2\ov M^2}\biggr)\Biggr]\;.\label{cjp}
\eea
Shifting an arbitrary parameter $M^2$ in Eq.~(\ref{cjp}) as 
$M^2 \rar M^2\exp[(11/3)-(49/54)\l/(4\p)^2+(563/1944)\l^2/(4\p)^4]$, 
we can readily see that leading order part of Eq.~(\ref{3p}) in $1/N$ agrees with
overlap part of leading-order approximation calculation of Coleman, Jackiw and 
Politzer \cite{cjp}. 

\acknowledgments               
The authors
are very grateful to Prof. Anatoly Kotikov at JINR, Dubna for his 
factorization check of the results [$J(a)$ to $L(d)$ in Eq.~(\ref{sum})].
A failure of factorization in $K(c)$ in the first version was recognized 
by him, thereby $K(c)$ was corrected in this new version. They are also
thankful to Prof. R. Jackiw for his interest in comparing our result of 
effective potential calculation with his earlier works \cite{jk,cjp}.       
This work was supported in part by Ministry of Education, Project number               
1998-015-D00073. J.~-M. C. was also supported in           
part by the Korea Science and Engineering Foundation, and in part by funds 
provided by the U.S. Department of Energyin part by funds provided by the 
U.S. Department of Energy (D.O.E.) under cooperative research agreement 
No.~DF-FC02-94ER40818.

\appendix            
\section*{One- and Two-Loop Integrals}          
\setcounter{equation}{0}           
We list one- and two-loop integrals which appear in our calculation of the        
three-loop integrals. The one-loop integrations are quite elementary and        
for the two-loop integrations one may refer to Ref.~\cite{2lp}. We need to        
calculate one-loop integrals up to the order of $\e$ except an overall        
multiplying factor $(\s^2/(4\p\m^2))^{-\e/2}$, two-loop integrals up to       
the order of $\e^0$ except an overall multiplying factor       
$(\s^2/(4\p\m^2))^{-\e}$, in conformity with our desired accuracy of       
three-loop integrals, i.e., the order of $\e^{-1}$ except an overall       
multiplying factor $(\s^2/(4\p\m^2))^{-3\e/2}$.             
      
\ni  \underline{One-loop integrals, $S_1$ to $S_4$}:             
\bea                 
S_1&\equiv&\int_k{1\ov k^2+\s^2}                 
={\s^2\ov (4\p)^2}\biggl({\s^2\ov 4\p\m^2}                 
\biggr)^{\!\!\!-\e/2}\G\biggl({\e\ov 2}-1\biggr)\;,\nn\\                 
S_2&\equiv&\int_k{1\ov k^2+\s^2/3}                 
={\s^2\ov 3(4\p)^2}\biggl({\s^2/3\ov 4\p\m^2}                 
\biggr)^{\!\!\!-\e/2}\G\biggl({\e\ov 2}-1\biggr)\;,\nn\\                 
S_3&\equiv&                 
\int_k{1\ov (k^2+\s^2)^2}={1\ov (4\p)^2}\biggl({\s^2\ov 4\p\m^2}                 
\biggr)^{\!\!\!-\e/2}\G\biggl({\e\ov 2}\biggr)\;,\nn\\                 
S_4&\equiv&                 
\int_k{1\ov (k^2+\s^2/3)^2}={1\ov (4\p)^2}\biggl({\s^2/3\ov 4\p\m^2}                 
\biggr)^{\!\!\!-\e/2}\G\biggl({\e\ov 2}\biggr)\;.  \label{uuu}              
\eea               
\ni  \underline{Two-loop integrals, $W_1$ to $W_8$}:          
\bea                 
W_1&\equiv&\int_{kp}{1\ov (p^2+\s^2)[(p+k)^2+\s^2]}                 
={\s^4\ov (4\p)^4}\biggl({\s^2\ov 4\p\m^2}                 
\biggr)^{\!\!\!-\e}\G^2\biggl({\e\ov 2}-1\biggr)\;,\nn\\                 
W_2&\equiv&\int_{kp}{1\ov (p^2+\s^2/3)                 
[(p+k)^2+\s^2/3]}={\s^4\ov 9(4\p)^4}\biggl({\s^2/3\ov 4\p\m^2}                 
\biggr)^{\!\!\!-\e}\G^2\biggl({\e\ov 2}-1\biggr)\;,\nn\\                 
W_3&\equiv&\int_{kp}{1\ov (p^2+\s^2)                 
[(p+k)^2+\s^2/3]}={\s^4\ov 3(4\p)^4}\biggl({\s^2\ov 4\p\m^2}                 
\biggr)^{\!\!\!-\e/2}\biggl({\s^2/3\ov 4\p\m^2}\biggr)^{\!\!\!-\e/2}                 
\G^2\biggl({\e\ov 2}-1\biggr)\;,\nn\\                 
W_4&\equiv&\int_{kp}{1\ov (k^2+\s^2)                 
(p^2+\s^2)[(p+k)^2+\s^2]}\nn\\                
&=&{\s^2\ov (4\p)^4}\biggl({\s^2\ov 4\p\m^2}\biggr)^{\!\!\!-\e}                 
{\G^2(1+\e/2)\ov (1-\e)(1-\e/2)}\biggl[-{6\ov \e^2}-3A+O(\e)           
\biggr]\;,\nn\\                 
W_5&\equiv&               
\int_{kp}{1\ov (k^2+\s^2)(p^2+\s^2/3)[(p+k)^2+\s^2/3]}\nn\\                 
&=&{\s^2\ov 3(4\p)^4}\biggl({\s^2/3\ov                  
4\p\m^2}\biggr)^{\!\!\!{-\e}}                 
{\G^2(1+\e/2)\ov (1-\e)(1-\e/2)}\biggl[-{10\ov \e^2}+{6\ov \e}\ln 3           
-{3\ov 2}                 
\ln^2 3-B+O(\e)\biggr]\;,\nn\\                 
W_6&\equiv&               
\int_{kp}{1\ov (k^2+\s^2)^2(p^2+\s^2)[(p+k)^2+\s^2]}\nn\\                 
&=&{1\ov (4\p)^4}\biggl({\s^2\ov 4\p\m^2}                 
\biggr)^{\!\!\!-\e}\biggl[{2\ov \e^2}+{1\ov \e}\biggl\{1-2\g\biggr\}+                 
{1\ov 2}-\g+\g^2+{\p^2\ov 12}+A+O(\e)\biggr]\;,\nn\\                 
W_7&\equiv&               
\int_{kp}{1\ov (k^2+\s^2/3)^2(p^2+\s^2/3)[(k+p)^2+\s^2]}\nn\\                 
&=&{1\ov (4\p)^4}\biggl({\s^2/3\ov 4\p\m^2}                 
\biggr)^{\!\!\!-\e}\biggl[{2\ov \e^2}+{1\ov \e}\biggl\{1-2\g\biggr\}+                 
{1\ov 2}-\g+\g^2+{\p^2\ov 12}+B+O(\e)\biggr]\;,\nn\\                 
W_8&\equiv&               
\int_{kp}{1\ov (k^2+\s^2)^2(p^2+\s^2/3)[(k+p)^2+\s^2/3]}\nn\\                 
&=&{1\ov (4\p)^4}\biggl({\s^2\ov 4\p\m^2}                 
\biggr)^{\!\!\!-\e}\biggl[{2\ov \e^2}+{1\ov \e}\biggl\{1-2\g\biggr\}+                 
{1\ov 2}-\g+\g^2+{\p^2\ov 12}+C+O(\e)\biggr]\;.\label{www}               
\eea                 
In the above equation, $\g$ is the usual Euler constant,              
$\g=0.5772156649\cdots$, and numerical values of the constants,              
$A$, $B$, and $C$ in Eq.~(\ref{www}) are                 
\bea              
A=f(1,1)=-1.1719536193\cdots\;,\nn\\                
B=f(1,3)=-2.3439072387\cdots\;,\nn\\                
C=f\Bigl({1\ov 3},{1\ov 3}\Bigr)=0.1778279325\cdots\;,  \label{abc}              
\eea               
where               
\beas               
&& f(a,b)\equiv\int_0^1dx\biggl[\int_0^{1-z}dy\biggl(-{\ln(1-y)\ov y}           
\biggr)                 
-{z\ln z\ov 1-z}\biggr]\;,~~~z={ax+b(1-x)\ov x(1-x)}\;.               
\eeas                        
These constants $A$, $B$, and $C$ can be analytically integrated \cite{kt}.
The results are expressed in terms of Clausen function:
\beas
A={B\ov 2}=-{3\ov 2}C-{3\ov 4}\ln^2 3=
-{2\ov \sqrt{3}}{\rm Cl}_2\Bigl({\p\ov 3}\Bigr)\;,
\eeas
where
\beas 
{\rm Cl}_2(\tt)=\int_0^\tt\ln[2\sin(\tt'/2)]d\tt'\;. 
\eeas


\begin{figure}          
  {\unitlength1cm           
   \epsfig{file=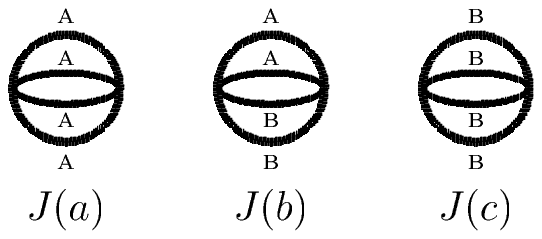,           
   bbllx=65pt,bblly=0pt,bburx=612pt,bbury=650pt,           
      rheight=2.6cm, rwidth=10cm,  clip=,angle=0} }             
\caption{}     
\end{figure}          
          
\begin{figure}          
  {\unitlength1cm           
   \epsfig{file=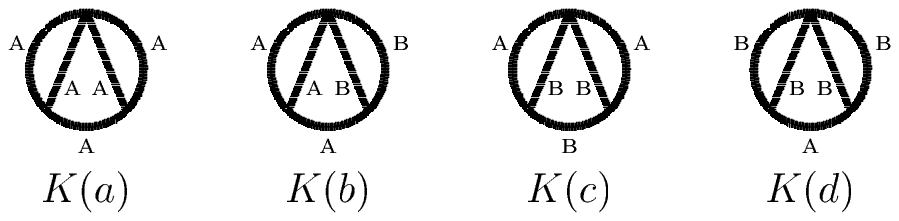,           
   bbllx=80pt,bblly=0pt,bburx=612pt,bbury=650pt,           
      rheight=2.6cm, rwidth=10cm,  clip=,angle=0} }             
\caption{}     
\end{figure}          
          
\begin{figure}          
  {\unitlength1cm           
   \epsfig{file=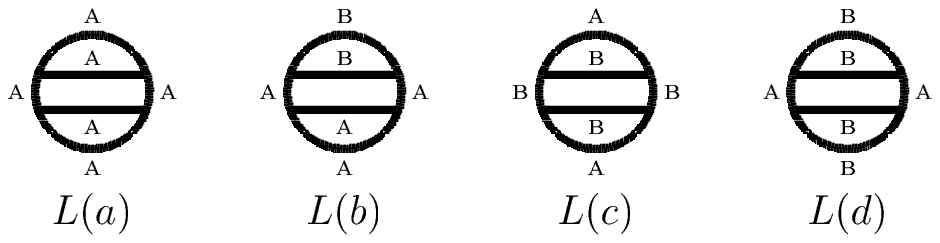,           
   bbllx=80pt,bblly=0pt,bburx=612pt,bbury=650pt,           
      rheight=2.6cm, rwidth=10cm,  clip=,angle=0} }             
\caption{}             
\end{figure}          
          
\begin{figure}          
  {\unitlength1cm           
   \epsfig{file=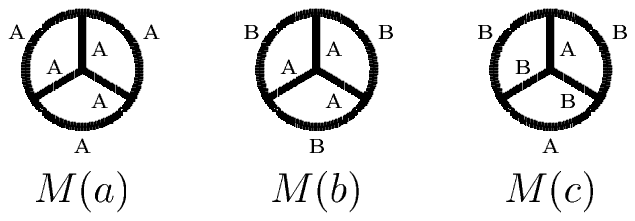,           
   bbllx=80pt,bblly=0pt,bburx=612pt,bbury=650pt,           
      rheight=2.5cm, rwidth=10cm,  clip=,angle=0} }             
\caption{}     
\end{figure}

\end{document}